\documentclass[12pt,letterpaper]{article}

\usepackage{color}
\usepackage{amsmath,amsthm,amssymb,amsfonts}
\usepackage{bm,epsfig,epsf,url,dsfont}
\usepackage{fullpage}
\usepackage[top=1in,bottom=1in,left=1in,right=1in]{geometry}
\usepackage{graphicx}
\usepackage{newtxmath}      % vvmathbb
\usepackage{hyperref,xcolor}
\hypersetup{
  colorlinks=true,
  linkcolor=blue,
  linktoc=all,
  citecolor=blue,
  urlcolor=blue,
  bookmarksnumbered=true
}
\usepackage{cleveref}

\newtheorem{theorem}{Theorem}

\newtheorem{definition}[theorem]{Definition}

\newtheorem{proposition}[theorem]{Proposition}

\newtheorem{lemma}[theorem]{Lemma}

\title{Smoothed Analysis of Online Metric Problems}
\author{
  Christian Coester\thanks{Department of Computer Science, University of Oxford. Email: christian.coester@cs.ox.ac.uk}
  \and
  Jack Umenberger\thanks{Department of Engineering Science, University of Oxford. Email: jack.umenberger@eng.ox.ac.uk}
}
\usepackage{bm,epsfig,epsf,url,dsfont}
\usepackage{newtxmath}
\usepackage{amsfonts}

\newcommand{\cO}{\mathcal{O}}
\newcommand{\E}{{\mathbb{E}}}

\newcommand{\R}{{\mathbb{R}}}

\DeclareMathOperator{\cost}{cost}
\DeclareMathOperator{\opt}{opt}
\DeclareMathOperator{\poly}{poly}
\DeclareMathOperator{\vol}{vol}
\DeclareMathOperator{\diam}{diam}

\newcommand{\Ind}{\vvmathbb 1}
\let\1\Ind

\begin{document}

\maketitle

\begin{abstract}
    We study three classical online problems -- $k$-server, $k$-taxi, and chasing size $k$ sets -- through a lens of smoothed analysis. Our setting allows request locations to be adversarial up to small perturbations, interpolating between worst-case and average-case models. Specifically, we show that if the metric space is contained in a ball in any normed space and requests are drawn from distributions whose density functions are upper bounded by $1/\sigma$ times the uniform density over the ball, then all three problems admit polylog$(k/\sigma)$-competitive algorithms. Our approach is simple: it reduces smoothed instances to fully adversarial instances on finite metrics and leverages existing algorithms in a black-box manner. We also provide a lower bound showing that no algorithm can achieve a competitive ratio sub-polylogarithmic in $k/\sigma$, matching our upper bounds up to the exponent of the polylogarithm. In contrast, the best known competitive ratios for these problems in the fully adversarial setting are $2k-1$, $\infty$ and $\Theta(k^2)$, respectively.
\end{abstract}

\section{Introduction}

In online algorithms, an algorithm has to make decisions online while the input is being revealed, without knowledge of the future input. Traditionally, the majority of research in the field has analyzed problems in the worst-case regime, in terms of the \emph{competitive ratio}, i.e., the largest possible ratio between the algorithm's performance and the optimum in hindsight. This perspective enables clean mathematical models and robust guarantees for algorithms that are satisfied regardless of the input they face. However, worst-case instances are often degenerate and unlikely to occur in practice. Tuning algorithms against such pathological scenarios can lead to worst-case-like performance even when actual inputs are much more favorable.

Moreover, the guarantees that can be achieved even under the most pessimistic assumptions are often weak by necessity. While tight competitive ratio bounds can appeal theoretically, algorithms are unlikely to be adopted in practice if the best promise is that they perform at most some large (e.g., polynomial or exponential) factor worse than optimal. Worse yet, for some problems such as the $k$-taxi problem, no finite bound on the competitive ratio is currently known for general metric spaces of infinitely many points, even if their diameter is bounded.

The research community generally understands the limitations of worst-case analysis, and there have been many demands for and an increased interest in beyond-worst-case analysis in recent years~\cite{Roughgarden20}.
A major success story of beyond-worst-case analysis has been Spielman and Teng's smoothed analysis~\cite{spielman2004smoothed}, spectacularly explaining why the simplex algorithm for linear problems performs well in practice despite its exponential worst-case running time. They show that perturbing any instance by small Gaussian noise yields an instance which the simplex algorithm solves in expected polynomial time. By varying the size of the random perturbations, smoothed analysis allows to interpolate between worst-case and average-case settings. Other examples of smoothed analysis of the running time of offline problems include $k$-means~\cite{ArthurV06,ArthurMR11}, TSP~\cite{EnglertRV14,MantheyV13,EnglertRV16}, local max-cut~\cite{AngelBPW17} and more general local search problems~\cite{GiannakopoulosG24}. We refer to~\cite{Roughgarden20} for more background.

In the context of online problems, prior smoothed analysis has focused mostly on regret minimization in online learning~\cite{rakhlin2011online,GuptaR17,Cohen-AddadK17,KannanMRWW18,RaghavanSVW18,HaghtalabRS20,HaghtalabRS24,BlockDGR22,HaghtalabHSY22,DurvasulaHZ23} and online discrepancy minimization~\cite{HaghtalabRS24,BansalJM0S22a,BansalJM0S22b}. In contrast, understanding is much more limited about the smoothed complexity of more dynamic online problems which are typically analyzed through their competitive ratio (rather than regret). The only examples we are aware of are the work of \cite{becchetti2006average}, which conduct smoothed competitive analysis of the multi-level feedback algorithm for non-clairvoyant scheduling, and \cite{schafer2005smoothedmts}, who analyze the work function algorithm for metrical task systems when cost function values are randomly perturbed.

\subparagraph{Smoothed Competitive Analysis in Metric Spaces.} Here, we initiate smoothed competitive analysis of online problems in metric spaces with randomly perturbed request locations. To do so, we study three classical online problems: namely, the \emph{$k$-server} problem~\cite{ManasseMS90}, its generalization the \emph{$k$-taxi} problem~\cite{FiatRR90,CoesterK19}, and \emph{chasing small sets}~\cite{ChrobakL91} (also known as metrical service systems, and equivalent to layered graph traversal~\cite{papadimitriou1991shortest,FiatFKRRV98}). In these problems, the goal is to minimize movement in a metric space while serving a sequence of requests. In the $k$-server problem, the algorithm controls $k$ servers and each request is a location that one of the servers must move to. The $k$-taxi problem generalizes this to the case where each request is a pair of locations, modelling the pick-up and drop-off location of a ride request. In chasing small sets, there is a single server, and each request is a set of up to $k$ points that the server must move to.

On general metric spaces, the best known competitive ratio is $\Theta(k)$ for the $k$-server problem~\cite{KoutsoupiasP95} and $\Theta(k^2)$ for chasing small sets~\cite{BubeckCR22}. For the $k$-taxi problem, the situation is even worse and it is not even known whether any finite competitive ratio can be achieved, even for $k=4$ servers on a line metric. Better bounds exist only in special cases such as \emph{finite} metrics of $n$ points or bounded aspect ratio $\Delta$, where the $k$-server problem admits competitive ratios of $O(\log n\log^2 k)$ and $O(\log\Delta\log^3k)$~\cite{bubeck2018kserver}, and the $k$-taxi problem an $O(\log^3\Delta\log^2(nk\Delta))$-competitive algorithm~\cite{GuptaKP24}. However, since $n$ and $\Delta$ can be infinite in general, these bounds provide no improvement even in seemingly simple metrics such as the line metric or $[0,1]^2$ with some norm.

We argue that smoothed analysis is a realistic model for these problems. For example, a customer's pick-up location in taxi apps is often determined through somewhat inaccurate GPS or by placing a pin on a map by hand, and even if an address is specified this usually does not correspond to an exact point but rather a small area on a map. Even in cases where exact precision in response to arriving requests is important, the way that inputs are generated is rarely fully adversarial and may be subject to noise. If smoothed analysis can drastically improve the competitive ratio of these problems, we believe that algorithms achieving these improved guarantees have a much higher chance of being adopted in practice than pessimistic worst-case methods.

We consider the setting where the metric space $M$ is a bounded diameter subset of some normed space (and therefore contained in some ball $B_M$). For each time step, an adversary can choose a distribution from which the next request is drawn, subject to the constraint that the density functions of points appearing in the requests are upper bounded by $1/\sigma$ times the density of the uniform distribution over $B_M$. Thus, as $\sigma\to 0$, this converges to the classical worst-case setting, whereas $\sigma=1$ means that request locations are drawn uniformly from $M=B_M$. Small values of $\sigma$ can model, for example, request locations that are slightly perturbed in a small ball around the worst-case locations.

For all three problems, we obtain $\text{polylog}(k/\sigma)$-competitive algorithms in this setting via simple reductions to the adversarial setting in finite metrics. Thus, for $\sigma$ polynomial in $k$ we obtain an exponential improvement over the known worst-case guarantees for $k$-server and chasing small sets, and an infinite improvement for the $k$-taxi problem. Our algorithms require no knowledge about $\sigma$ (except for the $k$-taxi problem, where a non-zero lower bound on $\sigma$ is requied), and beyond these smoothed bounds they still also achieve the known worst-case guarantees (where they exist, i.e., for $k$-server and chasing small sets).

\subparagraph{Further Related Work.} The $k$-server problem and an easier version of the $k$-taxi problem was previously studied under a stochastic setting different from smoothed analysis by Dehgani et al.~\cite{DehghaniEHLS17}. They consider the case where each request is drawn independently from some distribution, and these distributions are known to the algorithm. For this setting, they give an approximation algorithm for the line and finite metrics that is competitive against the best~\emph{online} algorithm for the instance. In contrast, we do not assume knowledge of the distributions, and compare against the best offline algorithm for the input.

Several online problems have been studied in a stochastic setting where requests are drawn i.i.d. from some distribution, see e.g.~\cite{GargGLS08,GrandoniGLMSS13,GuptaGPW19}. This differs from our setting, as we do not require distributions to be identical or independent.

Another beyond-worst-case model that has received significant interest in recent years is the learning-augmented paradigm~\cite{LykourisV21,MitzenmacherV22}, where an algorithm's input is augmented with additional predictions about future input or best behavior.

The $k$-server problem, introduced in~\cite{ManasseMS90}, is considered one of the most fundamental problems in online algorithms and is often called the ``holy grail of competitive analysis''. For deterministic algorithms the $\Theta(k)$ bound is known to be tight~\cite{ManasseMS90,KoutsoupiasP95} up to a constant factor, and the $k$-server conjecture states that it is exactly $k$. For randomized algorithms, the aforementioned improvements for the special cases of finite metrics and bounded aspect ratios are known~\cite{bubeck2018kserver}. The randomized $k$-server conjecture, positing an $O(\log k)$-competitive algorithm for general metrics, was recently refuted in~\cite{bubeck2023false}. The $k$-taxi problem is known to be at least exponentially harder than the $k$-server problem at least for deterministic algorithms, with an $\Omega(2^k)$ lower bound and no known upper bound for general metrics~\cite{CoesterK19}. For chasing small sets, the deterministic competitive ratio is known to lie in $O(k2^k)\cap\Omega(2^k)$~\cite{FiatFKRRV98,Burley96}, and the randomized competitive ratio is $\Theta(k^2)$~\cite{BubeckCR22,bubeck2023false}.

Chasing small sets is a special case of the metrical task systems (MTS) problem \cite{borodin1992mts}.
In MTS, each request consists of a cost function, and the algorithm which moves the single server incurs both the `service cost' as well as the movement cost. Chasing small sets corresponds to MTS with request cost functions restricted to set-membership indicator functions. The $k$-server problem is also a special case of MTS, though not in the same metric space but rather the metric space of configurations.

\subsection{Preliminaries}
All the problems we consider are defined over a metric space $(M,d)$. The input revealed online is a sequence of requests $r_1,r_2,\dots,r_T$, each of which must be served immediately as specified below.

\subparagraph{$k$-Server.}
The algorithm controls the movement of $k$ servers located at points in $M$. Each request is a point $r_t\in M$, and must be served by moving one of the servers to $r_t$. The cost of the algorithm is defined as the total distance travelled by all servers.

\subparagraph{$k$-Taxi.}
The algorithm controls the movement of $k$ taxis located at points in $M$. Each request is a pair of points $r_t=(a_t,b_t)\in M^2$, representing the start and destination of a ride request. To serve it, the algorithm must choose a taxi and move it first to $a_t$ and then to $b_t$. The cost is defined as the total distance of ``empty runs'' (i.e., distance travelled except for the travels from $a_t$ to $b_t$).

\subparagraph{Chasing Small Sets.}
The algorithm controls the movement of a single server located in $M$. Each request is a subset $r_t\subset M$ of cardinality at most $k$. To serve it, the algorithm must move the server to one of the points in $r_t$. The cost incurred by the algorithm is the total distance travelled.

\subparagraph{Smooth Instances.}
Throughout this paper, we assume that the metric space $M$ is a subset of $\R^m$ and the metric $d$ is induced by a norm, i.e., $d(x,y)=\|x-y\|$ for all $x,y\in M$, where $\|\cdot\|$ is an arbitrary norm.  We assume that $M$ has bounded diameter, and denote by $B_M$ a (closed) ball of smallest radius containing $M$, and by $R_M$ its radius.

We consider the variant of the problems described above where each request $r_t$ is drawn from some probability distribution $D_t$. These distributions need not be independent, i.e., the distribution $D_t$ (and whether it exists or the sequence stops earlier) can be chosen by the adversary after the realizations of $r_1,\dots,r_{t-1}$ have been determined (however, without knowledge of any random choices made by the online algorithm). For the $k$-server problem, $D_t$ is a distribution over $M$, for the $k$-taxi over $M^2$ and for chasing small sets over $M^{k_t}$ for some $k_t\le k$. We denote by $D_{ti}$ the corresponding marginal distributions, where $i=1$ for $k$-server, $i\in\{1,2\}$ for $k$-taxi and $i\in[k_t]$ for chasing small sets.

In the absence of any restrictions on the probability distribution, this problem setting is equivalent to the familiar worst-case setting, as the adversary can choose a (degenerate) distribution that places all probability mass on the worst-case request sequence. Thus, we impose the following smoothness assumption:

\begin{definition}\label{def:smooth}
    Let $\sigma\in(0,1]$. An instance is $\sigma$-smooth if for each marginal distribution $D_{ti}$, the probability of any measurable set $S\subseteq M$ is at most a factor $1/\sigma$ larger than under the uniform distribution over $B_M$.
\end{definition}
Equivalently, this means that the density function of each marginal distribution is upper bounded by $1/(\sigma\cdot\vol(B_M))$. The case $\sigma=1$ corresponds to uniformly distributed requests, whereas $\sigma\to0$ captures the classical worst-case setting.

\subparagraph{Competitive Ratio in the Smoothed Model.} For a (possibly randomized) algorithm $A$, we denote its expected cost on an instance $I$ by $\cost(A,I)$. Denote by $\opt$ the offline algorithm that achieves the lowest possible cost on any instance. In the classical (worst-case) setting, an online algorithm is said to be $c$-competitive if
\begin{align}
    \cost(A,I)\le c\cdot\cost(\opt,I) + \alpha,\label{eq:compRatio}
\end{align}
for all possible instances $I$, where $\alpha$ is a constant independent of $I$. We say that $A$ is $c$-competitive on $\sigma$-smooth instances if inequality \eqref{eq:compRatio} holds when $I$ is sampled as a $\sigma$-smooth instance and both sides of the inequality are replaced by their expectation over the randomness in $I$. We allow $\opt$ to serve the actual realization of the instance optimally, i.e., knowing the outcome of the random sampling process of $I$ in advance. Of course, any competitive algorithm in this setting is also competitive if $\opt$ has to make its choices before knowing the outcomes of the future sampling process of $I$.

We will usually drop $I$ from the notation when it is understood from the context.

\subparagraph{$\boldsymbol{\eta}$-Nets.} An \emph{$\eta$-net} of a metric space $M$ is a subset $N\subseteq M$ satisfying the following properties:
\begin{itemize}
    \item $N$ is $\eta$-dense in $M$: For all $x\in M$ there exists $y\in N$ such that $d(x,y)\le \eta$.
    \item $N$ is $\eta$-separated: For all $x,y\in N$ it holds that $d(x,y)>\eta$.
\end{itemize}

\subparagraph{Aspect Ratio.} The aspect ratio of a metric space is the ratio between the largest and smallest non-zero distance.

\subsection{Our Results}
We obtain polylog$(k/\sigma)$-competitive algorithms for all three problems we consider, as summarized in the following theorems. By known algorithm combination techniques, our algorithms for $k$-server and chasing small sets still also achieve the same robustness guarantees as classical worst-case algorithms.

\begin{theorem}[$k$-Server]\label{thm:kserver}
    There exists an $\cO(\min\{k,\log(k/\sigma)\cdot \log^2k\})$-competitive randomized algorithm for $k$-server on $\sigma$-smooth instances, even if $\sigma$ is unknown.
\end{theorem}

\begin{theorem}[$k$-Taxi]\label{thm:ktaxi}
    There exists an $\cO(\log^5(k/\sigma))$-competitive randomized algorithm for $k$-taxi on $\sigma$-smooth instances, provided a non-zero lower bound on $\sigma$ is known.
\end{theorem}

\begin{theorem}[Chasing Small Sets]\label{thm:ChasingSmallSets}
    There exists an $\cO(\min\{k^2, \log^2(k/\sigma)\})$-competitive randomized algorithm for chasing small sets on $\sigma$-smooth instances, even if $\sigma$ is unknown.
\end{theorem}

We complement our upper bounds with the following lower bound, ruling out algorithms with a sub-polylogarithmic dependence on $k/\sigma$.

\begin{theorem}[Lower Bound]\label{thm:LB}
For any $c<1/2$, there is no $\cO(\log^c(k/\sigma))$-competitive randomized algorithm for general $\sigma$-smooth instances of $k$-server, $k$-taxi, or chasing small sets when the metric space is a ball in $O(\log k)$-dimensional $\ell_\infty$-space, even if all distributions $D_t$ are identical and independent.
\end{theorem}

\subsection{Organization} All our algorithms are based on the same framework, as explained in the next section. The upper bound proofs are completed by separate arguments for bounding the offline cost for the three problems in Section~\ref{sec:offlineCost}. The polylogarithmic lower bound (Theorem~\ref{thm:LB}) is proved in Section~\ref{sec:LB}.

\section{Basic Construction}\label{sec:idea}

To describe our algorithm, let us assume initially that $\sigma$ is known. Later we will show how to drop this assumption.

For some suitable $\eta>0$, we choose a finite $\eta$-net $N$ of the metric space $M$. For each $x\in M$, we pick some point $y\in N$ satisfying $d(x,y)\le \eta$ and call it the \emph{projection of $x$ onto $N$}, denoted by $\pi(x)=y$ (making an arbitrary choice in case there are several options). Similarly, for a request $r_t$ we denote by $\pi(r_t)$ the projection of $r_t$, where each point in $r_t$ is replaced by its projection.
As the request sequence $r_1,r_2,\dots,r_T$ in $M$ is revealed, we simulate some online algorithm $A_N$ that operates on finite metrics to serve the request sequence $\pi(r_1),\pi(r_2),\dots,\pi(r_T)$ in metric space $(N,d)$.
Our online algorithm $A_M$ for the original problem on $M$ follows the configurations of $A_N$ at all times, except for an additional movement from $\pi(r_t)$ to $r_t$ and back at time $t$, to make sure $A_M$ serves the original request sequence. (For example, for a taxi request $(a_t,b_t)$, algorithm $A_M$ first moves the same taxi as $A_N$ to the point $\pi(a_t)$, then pays at most $\eta$ to move from $\pi(a_t)$ to $a_t$; from here, it serves the ride request, which changes the taxi's location from $a_t$ to $b_t$ (at no cost); finally, it moves from $b_t$ to $\pi(b_t)$ for another cost of at most $\eta$, restoring the invariant that $A_M$ is in the same configuration as $A_N$.) Thus, the cost of $A_M$ exceeds the cost $A_N$ by at most $2\eta$ per request, and the total cost of $A_M$ is at most $2\eta \cdot T+\eta \cdot k$ larger for a sequence of $T$ requests, with the additional cost $\eta k$ for projecting the initial configuration onto $N$. (In the case of chasing small sets, the $\eta\cdot k$ term can also be replaced by $\eta$.)

\subsection{Reduction to Worst-Case Setting}

Our analysis is based on the observation that the smoothing model forces the offline optimal algorithm to incur a sufficiently large movement cost.
In particular, we have:
\begin{proposition}\label{prop:opt_lower_bound}
    For $\sigma$-smooth instances of $k$-server, $k$-taxi, and chasing small sets, the offline optimal cost is at least $\Omega\left(R_M\cdot\left(\frac{\sigma}{\poly(k)}\right)^{1/m}\right)$ amortized per request in expectation, where $\poly(k)=O(k)$ for $k$-server and $k$-taxi and $\poly(k)=O(k^2)$ for chasing small sets.
\end{proposition}
More precise statements for the respective problems along with a proof of \Cref{prop:opt_lower_bound} are given in \Cref{sec:offlineCost}.

If $\eta$ is within a constant factor of the bound in \Cref{prop:opt_lower_bound} and algorithm $A_N$ is $c_N$-competitive for the worst-case (non-smoothed) problem in $N$, then we can show that algorithm $A_M$ described above is $\cO(c_N)$-competitive for the smoothed problem in $M$:

\begin{lemma}\label{lem:grid_to_continuous}    
    If $A_N$ is $c_N$-competitive in metric space $(N,d)$, then algorithm $A_M$ is $\cO(c_N)$-competitive on instances in $M$ whose optimal offline cost is at least $\Omega(\eta)$ amortized per request.
\end{lemma}
\begin{proof}
    Denote by $\opt_M$ and $\opt_N$, respectively, an optimal offline algorithm for the instance in $M$ and the projected instance in $N$. By assumption
    \begin{equation}\label{eq:gridalg_competitive}
        \cost(A_N)\le c_N \cdot \cost(\opt_N) + \alpha
    \end{equation}
    for some constant $\alpha$.
    Denote by $T$ the number of requests and by $\E[T]$ its expectation\footnote{$T$ may be random as the adversary can choose the stopping point depending on the realizations of earlier requests.}.
    As argued above,
    \begin{equation}\label{eq:alg_gridalg_comparison}
        \cost(A_M) \leq \cost(A_N) + 2\eta\cdot\E[T] + \eta \cdot k,
    \end{equation}
    with the additional cost due to movement between $\pi(r_t)$ and $r_t$ at each time. 
    By a similar argument,
    \begin{equation}\label{eq:gridopt_opt_comparison}
        \cost(\opt_N) \leq \cost(\opt_M) + 2\eta\cdot\E[T], 
    \end{equation}   
    as a possible offline algorithm to service the requests $\pi(r_1),\pi(r_2),\dots,\pi(r_T)$ in $N$ is to always be in the configuration of $\opt_M$ projected onto $N$, incurring at most an additional cost $2\eta$ per request by the triangle inequality (noting that, without loss of generality, $\opt_M$ moves at most one server/taxi per time step).
    Combining inequalities \eqref{eq:gridalg_competitive}, \eqref{eq:alg_gridalg_comparison} and \eqref{eq:gridopt_opt_comparison} gives
    \begin{equation*}
        \cost(A_M) \leq c_N \cdot \cost(\opt_M) + \left(c_N+1\right) \cdot 2\eta\cdot\E[T] + \eta\cdot k + \alpha.
    \end{equation*}
    Next, we make use of the assumption that the optimal offline cost is at least $\Omega(\eta)$ amortized per request in expectation, i.e., $\cost(\opt_M)\ge\Omega(\eta)\cdot\E[T] - \beta$ for some constant $\beta$. Substituting this into the previous inequality gives
    \begin{equation*}
        \cost(A_M) \leq \cO(c_N) \cdot \cost(\opt_M) + \eta\cdot k + \alpha + \cO(\beta\cdot c_N),
    \end{equation*}
    and the lemma follows since the additive term $\eta\cdot k + \alpha + \cO(\beta\cdot c_N)$ is a constant independent of the request sequence.
\end{proof}

\subsection{Size of the Net $N$}

\Cref{lem:grid_to_continuous} allows us to reduce smoothed instances on $M$ to non-smooth instances on $N$, provided the offline optimal cost is at least $\Omega(\eta)$ per request. For each of the problems we consider, by known results for finite metric spaces, the competitive ratio $c_N$ can be expressed as a function of the number of points in $N$ and possibly the aspect ratio $\Delta$ of $N$. Specifically, for $k$-server, the (best known) competitive ratio in $n$-point metric spaces is $\cO(\log^2 k\cdot\log n)$ \cite{bubeck2018kserver,BuchbinderGMN19}, for $k$-taxi it is $\cO(\log^3\Delta\cdot\log^2(nk\Delta))$~\cite{GuptaKP24}, and for chasing small sets (which is a special case of metrical task systems) it is $O(\log^2 n)$~\cite{bubeck2021unfair,CoesterL22}.

Thus, it remains to choose $\eta$ (i) large enough to obtain an appropriate value for the cardinality and aspect ratio of $N$ and (ii) small enough to ensure the lower bound in \Cref{prop:opt_lower_bound} becomes $\Omega(\eta)$.

The following lemma provides the desired relationship between the parameter $\eta$ and the number of points in $N$.

\begin{lemma}\label{lem:sizeN}
    For each $\eta\in(0,R_M]$, there exists an $\eta$-net $N$ of $M$ of size $|N|\le \left(\frac{3R_M}{\eta}\right)^m$.
\end{lemma}
\begin{proof}
    The proof is similar to \cite[Claim 2.6]{FandinaBartal18}. We initialize $N$ as a singleton set containing an arbitrary point from $M$; then as long as there exists a point $x\in M$ such that $d(x,y)>\eta$ for all $y\in N$, we add $x$ to $N$. By construction, $N$ is an $\eta$-net. Moreover, the balls of radius $\eta/2$ centered at the points in $N$ are disjoint and contained in a ball of radius $R_M+\eta/2$ (namely, the ball obtained by extending the radius of $B_M$ by $\eta/2$). The ratio between the volume of a ball of radius $\eta/2$ and the ball of radius $R_M+\eta/2$ is $\left(\frac{\eta/2}{R_M+\eta/2}\right)^m$. Thus, there can be at most $\left(\frac{R_M+\eta/2}{\eta/2}\right)^m \le \left(\frac{3R_M}{\eta}\right)^m$ disjoint balls.
\end{proof}

\subsection{Putting it all Together}

We now combine the above statements to prove the following statement.

\begin{proposition}
    For known $\sigma$, there exists a randomized online algorithm whose competitive ratio on $\sigma$-smooth instances is $\cO(\log(k/\sigma)\cdot\log^2 k)$ for $k$-server, $\cO(\log^5(k/\sigma))$ for $k$-taxi and $\cO(\log^2(k/\sigma))$ for chasing small sets.
\end{proposition}
\begin{proof}
    Let $\eta=3R_M\cdot\left(\frac{\sigma}{\poly(k)}\right)^{1/m}$, where $\poly(k)$ is the function from \Cref{prop:opt_lower_bound}. Then $A_M$ is $\cO(c_N)$-competitive by \Cref{prop:opt_lower_bound} and Lemma~\ref{lem:grid_to_continuous}. If $\eta > R_M$, then $N$ can be chosen as the singleton containing only the center of the ball $B_M$, which trivially yields $c_N=1$ and the theorems. Otherwise, by Lemma~\ref{lem:sizeN} the size of the $\eta$-net $N$ is
    \begin{align*}
        n:=|N|\le \frac{\poly(k)}{\sigma},
    \end{align*}
    and since the diameter of $M$ is at most $2R_M$, the aspect ratio of $N$ is
    \begin{align*}
        \Delta\le \frac{2R_M}{\eta} = \frac{2}{3}\left(\frac{\poly(k)}{\sigma}\right)^{1/m}.
    \end{align*}
    
    Thus, since $c_N=\cO(\log^2 k\cdot\log n)$ for $k$-server~\cite{bubeck2018kserver,BuchbinderGMN19}, $c_N=\cO(\log^3\Delta\cdot\log^2(nk\Delta))$ for $k$-taxi~\cite{GuptaKP24}, and $c_N=O(\log^2 n)$ for metrical task system~\cite{bubeck2021unfair,CoesterL22}, and chasing small sets is a special case of metrical task systems (in the same metric space), the result follows.
\end{proof}

We note that the competitive ratio for $k$-taxi could actually be improved by a factor $1/\min\{m,\log(k/\sigma)\}^3$. (Note that the $\log\Delta$ term stands for a quantity that is at least 1, so the improvement is not greater.) This would typically only yield a constant factor improvement though, considering practical applications of the $k$-taxi problem usually having a small dimension of at most $3$.

\subsection{Unknown $\sigma$ and Robustness}
If $\sigma$ is unknown, and to achieve the robustness bounds in our main theorems, the idea is to simulate in parallel several instances of the above algorithm for different values of $\sigma$ as well as a worst-case algorithm, and combine them using the technique of Blum and Burch~\cite{BlumB00}.

\begin{theorem}[Blum and Burch~\cite{BlumB00}]\label{thm:BB}
    Given $\epsilon>0$ and $\ell$ online algorithms $A_1,\dots,A_\ell$ for a problem that can be formulated as a metrical task system of diameter $\diam$, there exists an online algorithm whose expected cost is at most
    \begin{align*}
        (1+\epsilon)\cdot \min_{i\in[\ell]}\cost(A_i) + \cO\left(\frac{\diam\cdot\log\ell}{\epsilon}\right).
    \end{align*}
\end{theorem}

\subparagraph{Chasing Small Sets.} Since chasing small sets in a metric space $(M,d)$ is a metrical task system in the same metric space, we can invoke the above theorem with $\epsilon=1$, $\ell=\lceil\log_2 k\rceil+1$ and using as $A_i$ for $i=1,\dots,\ell-1$ our algorithm with smoothness parameter $\sigma_i=2^{-2^i}$, and as $A_{\ell}$ the $\Theta(k^2)$-competitive algorithm from~\cite{BubeckCR22} for chasing small sets in the fully adversarial setting.

Since $\diam\le2R_M$, the additive term in Theorem~\ref{thm:BB} is a constant independent of the request sequence. Thus, the resulting algorithm achieves, up to a factor $1+\epsilon\le2$, the same competitive ratio as the best of the algorithms $A_i$. If the true smoothness parameter $\sigma<2^{-(\ell-1)}\le 2^{-k}$, then $\cO(\min\{k^2, \log^2(k/\sigma)\})$-competitiveness follows since the minimum is attained by $k^2$. Otherwise, there exists $i$ such that $\sigma_i\le\sigma\le\sqrt{\sigma_i}$. Since any $\sigma$-smooth instance is also $\sigma_i$-smooth, we achieve competitive ratio $\cO(\log^2(k/\sigma_i)) = \cO(\log^2(k/\sigma))$. This yields Theorem~\ref{thm:ChasingSmallSets}.

\subparagraph{$k$-Server.} For the $k$-server problem (Theorem~\ref{thm:kserver}), the same argument works since $k$-server can be modelled as a metrical task system in the space of server configurations (whose diameter is $k$ times the diameter of the original metric space). The only difference is that the algorithm $A_\ell$ for general metrics in the fully adversarial setting has competitive ratio $\Theta(k)$~\cite{KoutsoupiasP95}.

\subparagraph{$k$-Taxi.} It is not known how to model the $k$-taxi problem as a metrical task system, due to configuration changes that incur no cost.\footnote{Though it belongs to the more general class of metrical task systems with transformation~\cite{BubeckBCS21}.} Nonetheless, the proof of Theorem~\ref{thm:BB} in~\cite{BlumB00} extends to this setting without any change, so we can still apply the same combination method also to the $k$-taxi problem. However, since no competitive algorithm for the $k$-taxi problem on general metrics in the fully adversarial setting is known, we require a lower bound on $\sigma$ to ensure that the number of combined algorithms is finite.

\section{Bounding Offline Cost}\label{sec:offlineCost}
\subsection{$k$-Server}
\begin{lemma}\label{lem:kserverOptLB}
    For $\sigma$-smooth instances of the $k$-server problem, the optimal offline algorithm pays an expected cost of at least
    \begin{align*}
        \cost(\opt_M)&\ge \frac{R_M}{8}\cdot\left(\frac{\sigma}{8k}\right)^{1/m}\cdot \E[T]-c,
    \end{align*}
    where $c>0$ is a constant independent of the request sequence.
\end{lemma}
\begin{proof}
    Let $\delta>0$. Consider a subsequence $r_{t+1},r_{t+2},\dots,r_{t+4k}$ of $4k$ consecutive requests. For $i=0,\dots,4k$, we denote by $S_i$ a $\delta$-separated subset of $\{r_{t+1},r_{t+2},\dots,r_{t+i}\}$, constructed as follows: $S_0=\emptyset$, and for $i\ge 1$,
    \begin{align*}
        S_i=\begin{cases}
            S_{i-1}\cup\{r_{t+i}\}\quad&\text{if $S_{i-1}\cup\{r_{t+i}\}$ is $\delta$-separated}\\
            S_{i-1}&\text{otherwise.}
        \end{cases}
    \end{align*}
    Note that $S_i=S_{i-1}$ if and only if $r_{t+i}$ belongs to a ball of radius $\delta$ centered at a point in $S_{i-1}$. The volume of each such ball is at most a $\left(\frac{\delta}{R_M}\right)^m$ fraction of the volume of $B_M$, and thus its probability under $D_{t+i}$ is at most $\left(\frac{\delta}{R_M}\right)^m/\sigma$. Hence, since there are at most $4k$ such balls of radius $\delta$, the probability that $S_i=S_{i-1}$ is bounded as
    \begin{align*}
        P\left(S_i = S_{i-1}\right) \le \frac{4k}{\sigma}\left(\frac{\delta}{R_M}\right)^m.
    \end{align*}
    We now choose $\delta = R_M\cdot\left(\frac{\sigma}{8k}\right)^{1/m}$, so that the right hand side becomes $1/2$. In other words, in each step, another point is added to $S_i$ with probability at least $1/2$. Thus, after $4k$ steps we have $|S_{4k}|\ge 2k$ with probability at least $1/2$. If this is the case, then any offline algorithm must pay at least $k\cdot\delta$ to serve these requests, regardless of the configuration it starts at: Indeed, at most $k$ of the requests in $S_{4k}$ can be served by a server that previously served no other request from $S_{4k}$. For the remaining $|S_{4k}|-k\ge k$ requests, the serving server must have moved at least distance $\delta$ since the previous time it served a request from $S_{4k}$, since $S_{4k}$ is $\delta$-separated. Thus, any subsequence of $4k$ consecutive requests contributes $k\cdot\delta/2$ to the expected optimal cost. Over a sequence of $T$ requests, this leads to an expected optimal cost of
    \begin{align*}
        \cost(\opt_M)&\ge \E\left[\left\lfloor\frac{T}{4k}\right\rfloor\right]\cdot \frac{k\cdot\delta}{2}\\
        &\ge \frac{\E[T]\cdot\delta}{8}-\frac{k\cdot\delta}{2}\\
        &= \frac{\E[T]\cdot R_M\cdot\left(\frac{\sigma}{8k}\right)^{1/m}}{8}-\frac{k\cdot R_M\cdot\left(\frac{\sigma}{8k}\right)^{1/m}}{2}
    \end{align*}
\end{proof}

\subsection{$k$-Taxi}
\begin{lemma}
    For $\sigma$-smooth instances of the $k$-taxi problem, the optimal offline algorithm pays an expected cost of at least
    \begin{align*}
        \cost(\opt_M)&\ge \frac{R_M}{8}\cdot\left(\frac{\sigma}{8k}\right)^{1/m}\cdot \E[T]-c,
    \end{align*}
    where $c>0$ is a constant independent of the request sequence.
\end{lemma}
\begin{proof}
    We proceed similarly to the proof of Lemma~\ref{lem:kserverOptLB}, except in the construction of $S_i$ we add a request $r_{t+i}=(a_{t+i},b_{t+i})$ if its start location $a_{t+i}$ is more than distance $\delta$ from the destinations of \emph{all} the requests $r_{t+1},\dots,r_{t+i-1}$:
    \begin{align*}
        S_i=\begin{cases}
            S_{i-1}\cup\{r_{t+i}\}\quad&\text{if $d(b_{t+j},a_{t+i})>\delta$ for all $j=1,\dots,i-1$}\\
            S_{i-1}&\text{otherwise.}
        \end{cases}
    \end{align*}
    Again,
    \begin{align*}
        P\left(S_i = S_{i-1}\right) \le \frac{4k}{\sigma}\left(\frac{\delta}{R_M}\right)^m
    \end{align*}
    as $S_i=S_{i-1}$ requires $a_{t+i}$ to fall in one of at most $4k$ balls of radius $\delta$. As before, the same choice of $\delta$ yields $|S_{4k}|\ge 2k$ with probability at least $1/2$. If this is the case, then any offline algorithm must pay at least $k\cdot\delta$ to serve these requests: At least $|S_{4k}|-k\ge k$ of the requests in $S_{4k}$ are served by a taxi that also served a previous request since time $t+1$, incurring cost at least $\delta$ by definition of $S_{4k}$. The lemma is then concluded in an identical fashion to the proof of Lemma~\ref{lem:kserverOptLB}.
\end{proof}

\subsection{Chasing Small Sets}
\begin{lemma}
    In $\sigma$-smoothed instances of chasing small sets, the optimal offline algorithm pays an expected cost of at least
    \begin{align*}
        \cost(\opt_M)&\ge \frac{R_M}{2}\cdot\left(\frac{\sigma}{2k^2}\right)^{1/m}\cdot \E[T].
    \end{align*}
\end{lemma}
\begin{proof}
Let $\delta>0$. Consider two consecutive request sets $r_{t-1}$ and $r_{t}$. The probability that the $i$th point in $r_{t-1}$ is within distance $\delta$ from the $j$th point in $r_{t}$ is at most $\frac{1}{\sigma}\left(\frac{\delta}{R_M}\right)^m$. Thus, by a union bound over the at most $k^2$ pairs of points in $r_{t-1}\times r_t$, the probability that some point in $r_{t-1}$ is within distance $\delta$ from some point in $r_{t}$ is at most $\frac{k^2}{\sigma}\left(\frac{\delta}{R_M}\right)^m$. Otherwise, with probability at least $1-\frac{k^2}{\sigma}\left(\frac{\delta}{R_M}\right)^m$, any offline algorithm must pay at least $\delta$ at time $t$ to move from $r_{t-1}$ to $r_t$. Choosing $\delta = R_M\cdot\left(\frac{\sigma}{2k^2}\right)^{1/m}$, we see that the probability is at least 1/2. Thus, the expected offline cost is at least $\frac{\delta}{2} = \frac{R_M}{2}\cdot\left(\frac{\sigma}{2k^2}\right)^{1/m}$ per time step.
\end{proof}

\section{Lower Bound}\label{sec:LB}
Complementing our upper bounds, we now prove a polylog$(k/\sigma)$ lower bound for the three problems, i.e., that no algorithm can achieve a competitive ratio of $\cO(\log^c(k/\sigma))$ for any $c<1/2$ (Theorem~\ref{thm:LB}).

Our proof is based on the classical $\Omega(\log k)$ lower bound for $k$-server and MTS on $(k+1)$-point uniform metrics, in which each request location is drawn uniformly at random from the $k+1$ points~\cite{borodin1992mts}. As $\sigma$-smooth instances require distributions with infinite support, we intend to simulate this by substituting each of the $k+1$ points by a small neighborhood in normed space. We describe the construction for $k$-server first, and then explain how to adapt it to $k$-taxi and chasing small sets.

\subsection{$k$-Server} We first recall the $\Omega(\log k)$ lower bound from~\cite{borodin1992mts}:\footnote{The presentation in~\cite{borodin1992mts} is for MTS. We describe it here in the language of $k$-server.} The underlying metric space is a uniform metric of $k+1$ points (i.e., all pairwise distances are $1$) and each request is drawn uniformly at random from the $k+1$ points. Thus, an online algorithm must move at each time step with probability $1/(k+1)$, paying $T/(k+1)$ in expectation for a request sequence of length $T$. In contrast, an offline algorithm can, whenever a request arrives at the (unique) location not covered by a server, take the server from the location whose next request is furthest in the future. Thus, it need not move again until each of the $k$ other locations is requested again at least once. By a standard coupon collector argument, it takes $\frac{k+1}{k}+\frac{k+1}{k-1}+\dots+\frac{k+1}{1}=\Theta(k\log k)$ steps in expectation for each of the $k$ other locations to be requested again, so the algorithm only needs to move roughly once every $\Theta(k\log k)$ steps. For a sequence of $T$ requests, this leads to an optimal offline cost of $\cO\left(\frac{T}{k\log k}+1\right)$ (see~\cite{borodin1992mts} for details). The ratio between the online cost $T/(k+1)$ and offline cost $\cO\left(\frac{T}{k\log k}+1\right)$ yields the known $\Omega(\log k)$ lower bound for uniform metrics.

To simulate this in the smoothed setting, let $m=\lceil\log_2(k+1)\rceil$ and take as metric space the hypercube $[0,1]^m$ with $\ell_\infty$ distance, which is a ball in $\ell_\infty^m$. Let $V$ be a fixed set of some $k+1$ vertices of this hypercube (i.e., with integer coordinates). Let $\epsilon=\frac{1
}{2k\log_2 k}\le \frac{1}{4}$. We say two points are \emph{near} each other if their distance is at most $\epsilon$. Otherwise, they are \emph{far}. Let $P$ be the set of points in $[0,1]^m$ that are near some vertex in $V$. Consider the instance where each request is drawn uniformly at random from $P$.

For the online algorithm, each request incurs cost at least $\Omega(1/k)$ in expectation, since with probability at least $\frac{1}{k+1}$ the request arrives near a vertex $v$ whereas all online algorithm servers are near vertices in $V\setminus \{v\}$. For the offline algorithm, if each request arrived at the nearby vertex in $V$ instead of its actual location in $P$, then the instance would be precisely the aforementioned instance on a $(k+1)$-point uniform metric, with an offline cost of $\cO\left(\frac{T}{k\log k}+1\right)$ for $T$ requests. The fact requests arrive \emph{near} vertices instead of \emph{at} vertices only increases the offline cost by at most $\epsilon$ per request, as the distance between any two points near the same vertex is at most $\epsilon$. By choice of $\epsilon$, this still leads to an offline cost of $\cO\left(\frac{T}{k\log k}+1\right)$, and a competitive ratio lower bound of $\Omega(\log k)$.

It remains to express the lower bound in terms of $k/\sigma$. The overall hypercube $[0,1]^m$ has a volume of $1$, whereas the set $P$ has a volume of $(k+1)\epsilon^m$. Thus, $\sigma=\frac{1}{(k+1)\epsilon^m}=2^{-\Theta(\log^2k)}$, and $k/\sigma=2^{\Theta(\log^2k)}$. Hence, the lower bound can be written as $\Omega(\log^{1/2}(k/\sigma))$.

\subsection{$k$-Taxi} The construction is similar, but requests need to be pairs of two points $(a,b)$. For each request, we sample $a$ as above, and if $a$ is near a vertex $v$, then we sample $b$ uniformly at random from the points near $v$ as well. This ensures that an offline algorithm can keep its taxis near $k$ distinct vertices. The only difference in calculations is that $\sigma$ is larger by a factor $k+1$ as request destinations are sampled from distributions supported on points near a single vertex, but the asymptotic bound $\sigma=2^{-\Theta(\log^2k)}$ is unaffected by this.

\subsection{Chasing Small Sets} We again use a similar construction. Each request set contains $k$ points in $P$, sampled uniformly at random subject to the constraint that the points are near $k$ distinct vertices of $V$. Thus, for each request set there is one vertex in $V$ that is far from it. 
The online cost is still $\Omega(1/k)$ per request in expectation, as with probability at least $1/(k+1)$ the request set contains no point near the algorithm's server. The offline cost is also bounded as before, as whenever it is far from all points in the request set, it can first move to the vertex for which it will happen furthest in the future that it is far from all points in the request set; additionally, it pays at most $\epsilon$ per request to move to a nearby point in the request set. The lower bound $\Omega(\log^{1/2}(k/\sigma))$ follows in the same way.

\subsubsection*{Acknowledgments}
C. Coester is funded by the European Union (ERC grant CCOO 10116513). Views and opinions expressed are however those of the author(s) only and do not necessarily reflect those of the European Union or the European Research Council Executive Agency. Neither the European Union nor the granting authority can be held responsible for them.

\bibliographystyle{plainurl}
\bibliography{references}

\end{document}